\documentclass[amsmath,amssymb,reprint]{revtex4-1}

\usepackage[utf8]{inputenc}
\usepackage[T1]{fontenc}
\usepackage{mathptmx}
\usepackage{verbatim}
\usepackage{fdsymbol}
\usepackage{graphicx}
\usepackage{sidecap}
\usepackage{placeins}
\usepackage{xcolor}
\usepackage{graphicx}
\usepackage{dcolumn}

\begin{document}

\title{2$\theta$-burster for rhythm-generating circuits}
\author{Aaron Kelley} 
\affiliation{Neuroscience Institute, Georgia State University, Atlanta, GA, USA}
\author{Andrey Shilnikov}
\affiliation{Neuroscience Institute and Department of Mathematics and Statistics,\\ Georgia State University, Atlanta, USA}

\begin{abstract}
We propose and demonstrate the use of a minimal $2\theta$ model for endogenous bursters coupled in 3-cell  neural circuits. This $2\theta$ model offers the benefit of simplicity of designing larger neural networks along with an acute reduction  on the computation cost. 
\end{abstract}

\date{\today}

\maketitle

\section{\label{sec:Introduction} Introduction}

Neural networks called Central Pattern Generators (CPGs)
\cite{KCF05,CTMKF07,Marder1994752,NSLGK012,CPG,Bal1988,Marder1996,KatzHooper07}.
produce and control a great variety of rhythmic  motor behaviors, including heartbeat, respiration, chewing, and locomotion.  Many anatomically and physiologically diverse CPG circuits involve a three-cell motifs including the spiny lobster pyloric network \cite{CPG}, the {\em Tritonia} swim circuit \cite{CTMKF07}, and the {\em Lymnaea} respiratory CPGs \cite{Marder1994752}. 
 Pairing experimental studies and modeling studies have proven to be key to unlock insights into operational and dynamical principles of CPGs \cite{Kopell26102004,Ma87,Kopell88,Canavier1994,SKM94,Dror1999,Prinz2003}. Although various circuits and models of specific CPGs have been developed, it still remains unclear how CPGs achieve the level of robustness and flexibility observed in nature.  It is not clear either what mechanisms a single motor system can use to generate multiple rhythms, i.e., whether CPGs use dedicated circuitry for each function, or whether the same circuitry is multi-functional and can govern several behaviors \cite{Kristan2008b,Briggman2008,Shilnikov2008b}.

\begin{figure*}[t!]
\includegraphics[width=1.2\columnwidth]{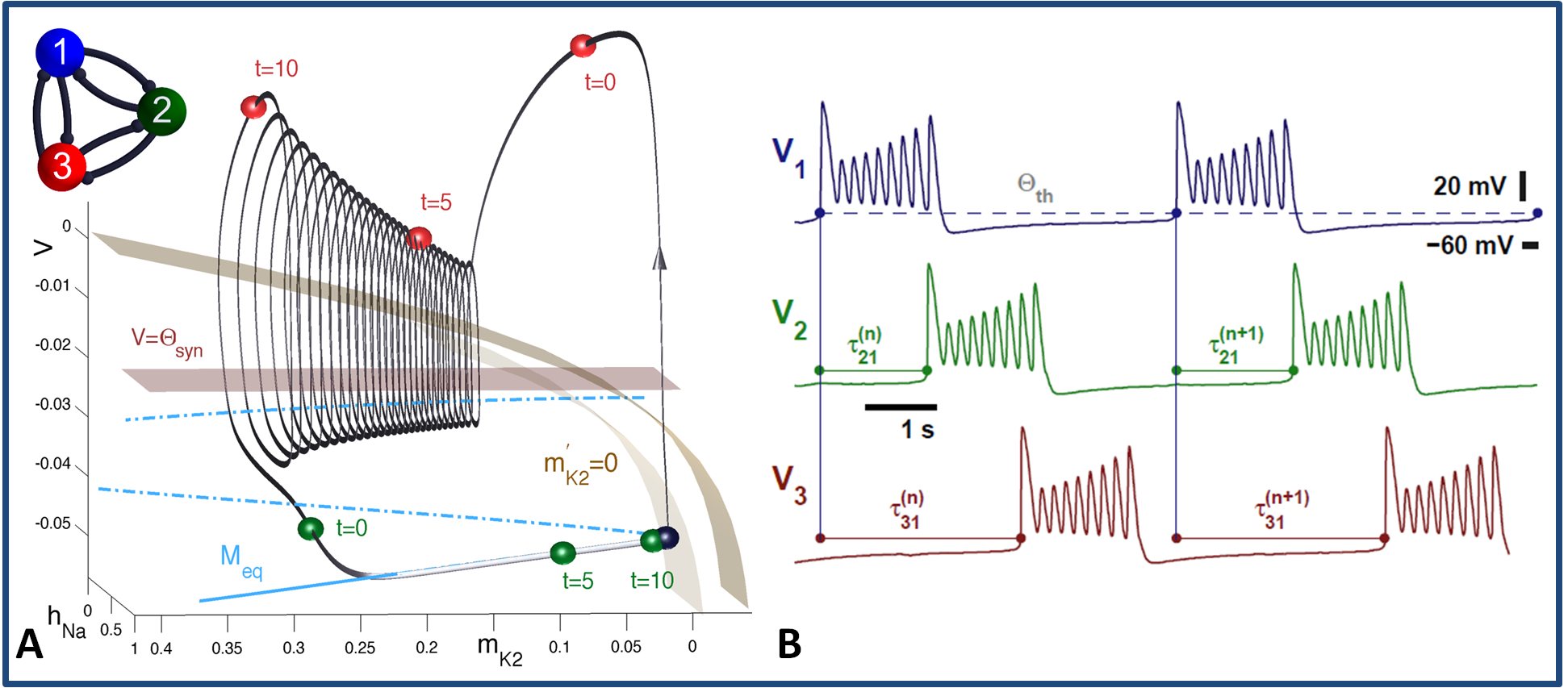}
\caption{(A) Snapshots of the current states (represented by the blue, green and red spheres) of three weakly-coupled endogenous bursters at $t=0$ and their further progressions at $t=10$, on the periodic orbit (grey) in the 3D phase space of the Hodgkin-Huxley type model of the leech heart interneuron \cite{rcd2004,Shilnikov2012}. The plane $V=\Theta_{\rm syn}$ represents the threshold for the chemical synapses, that divides the active ``on'' phase (above it) and the inactive ``off'' phase; here, the active red cell inhibits the quiescent  green and blue ones. (B) Burst initiations in successive voltage traces generated by the 3-cell neural network allow us to define the relative delays $\tau_{i1}$'s and hence, the phase lags (given by Eqs.~(\ref{phaselageq})) between its constituent bursters; see further details in \cite{Wojcik2011a,Wojcik2014}.}
\end{figure*}

This paper capitalizes on our original work and the well-established principles in the characterization of 3-cell circuits made of HH-type neurons \cite{prl08,Shilnikov2008b,pre2012,jalil2013} and the Fitzhugh-Nagumo-like neurons~\cite{schwabedal2016qualitative}.  We use the bottom-up approach to showcase the universality of rhythm-generation principles in 3-cell circuits regardless of the model selected, which can be a Hodgkin-Huxley (HH) type model of the leech heart interneuron \cite{rcd2004,Shilnikov2012},  the the generalized Fitzhugh-Nagumo (gFN) model of neurons~\cite{labpaper}, and the minimal 2$\theta$ bursting neuron, provided of course that all three models meet some simple and generic criteria.

\section{Return maps for phase lags} 

 We developed a computational toolkit for oscillatory networks that reduces the problem of the occurrence of bursting and spiking rhythms generated by a CPG network to the bifurcation analysis of attractors in the corresponding Poincar\'e return maps for the phase lags between oscillatory neurons. The structure of the phase space of the map is an individual signature of the CPG as it discloses all characteristics of the functional space of the network.  Recurrence of rhythms generated by the CPG (represented by a system of coupled Hodgkin-Huxley type neurons \cite{Shilnikov2012}) lets us employ Poincar\'e return maps defined for phase lags between spike/burst initiations in the constituent neurons  \cite{Wojcik2011a,Wojcik2014} as illustrated in Fig.~1,2 and 4. With such return maps, we can predict and identify the set of robust outcomes in a CPG with mixed, inhibitory/excretory and electrical synapses, which are differentiated by phase-locked or periodically varying lags corresponding, respectively, to stable fixed points and invariant circles of the return map.    
 
 \begin{figure}[b]
\includegraphics[width=0.99\columnwidth]{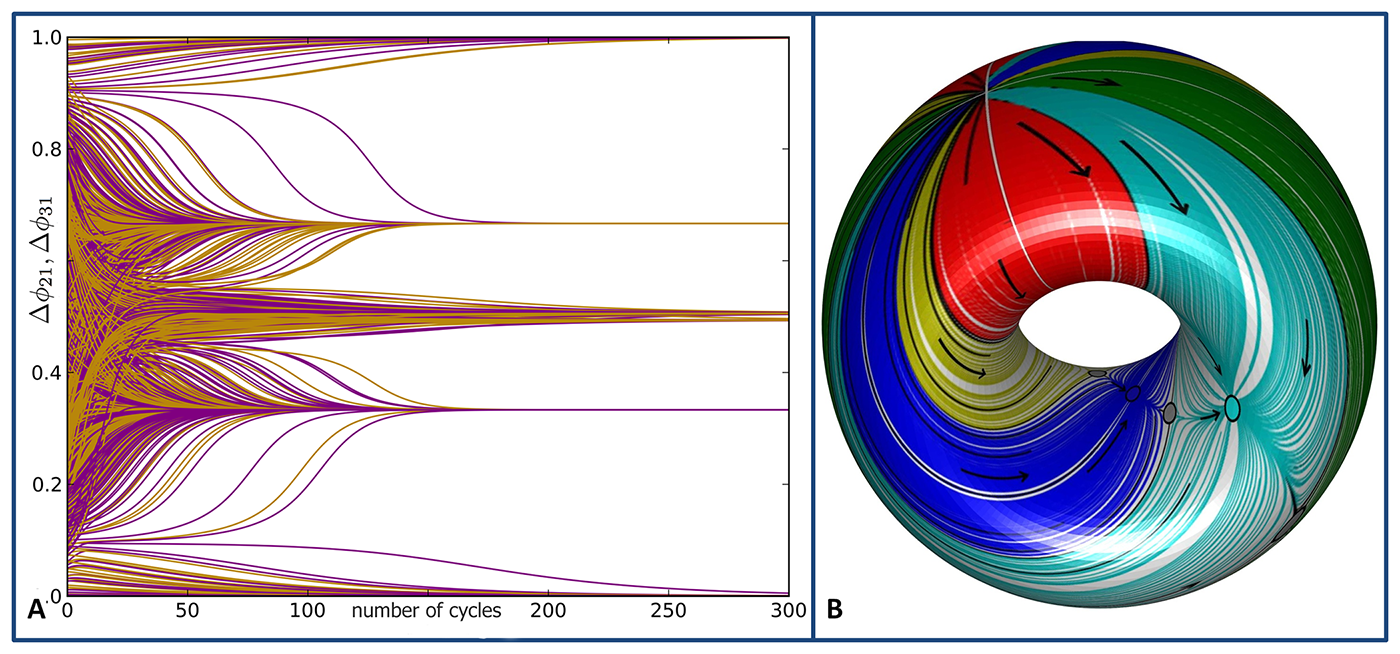}
\caption{(A) Slow exponential convergence of initial states of $\Delta \phi_{21}$ (yellow curves) and $\Delta \phi_{31}$ (purple curves) to four phase-locked states: $ \{0 \equiv 1,\, \frac{1}{3},\, \frac{1}{2}, \frac{2}{3}\}$, in the inhibitory 3-cell motif~(4) with weak coupling $\beta$ = 0.003. (B) Poincar\'e return map defined on a unit 2D torus, ${\mathbb T}^2 = {\mathbb S}^1 \otimes {\mathbb S}^1$ of two phase-lags, showing color-coded attraction basins of several fixed points (solid dots of same colors) corresponding to the phase-locked rhythms by the 3-cell motif. A flatten torus is shown in Fig.~3A.}
\end{figure}

 Let us introduce a 3-cell network (Fig.~1A) made of weakly coupled HH-like bursters; see the equations in the Appendix below. Here, ``weakly'' means that coupling cannot quite  disturb the shape of the stable bursting orbit in the 3D phase space of the individual HH-model (Fig.~1A).  Weak interactions, inhibitory (mainly repulsing) and excitatory/gap-junction (manly attracting)   can only affect the phases of the periodically varying states of the neurons, represented by the color-coded spheres, blue/green/red for cells 1/2/3, on the bursting orbit in the 3D phase space of the given interneuron model. As such weak-coupling can only gently alter the phase-differences or phase-lags between the coupled neurons (Fig.~2A). Being inspired by neuro-physiological recordings performed on various rhythmic CPGs, we employ only voltage traces generated by such networks to examine the time delays, $\tau_{21}$ and $\tau_{31}$  between the burst upstrokes on each cycle in the reference/blue cell 1 and in cells 2 (green) and 3 (red). In what follows, we will show that like the biologically plausible HH-type networks, 3-cell circuits of coupled 2$\theta$-bursters can stable produce similar phase-locked rhythms. They include, but not limited, peristaltic patterns or traveling waves, in which the cells burst sequentially one after the other  (see Figs.~1 and 3C/E), as well as the so-called pacemaker rhythms, in which one cell effectively inhibits and bursts in anti-phase with the other two  bursting synchronously (Fig.~3B/D). The symmetric connectivity implies such 3-cell networks can produce  multiple rhythms due to cyclic permutations of the constituent cells (see Fig.~3 below). To analyze the existence and the stability of various recurrent rhythms produced by such networks, we employ our previously developed approach using Poincar\'{e} return maps for phase-legs between constituent neurons. We introduce phase-lags  defined at specific events in time when the voltage in cells reaches some threshold value this signaling the burst initiation (see Fig.~1B). The phase lag $\Delta\phi^{(n)}_{1j}$ is then defined by a delay between $n$-th burst initiations in in the given cell and  the reference cell 1, normalized over the bursting period: 
\begin{equation}
\begin{array}{rcl}
\Delta \phi_{12}^{(n)} &=& \frac{t_{2}^{(n)}-t_{1}^{(n)}}{t^{(n+1)}_1-t^{(n)}_1}~,\\
\Delta \phi_{13}^{(n)} &=& \frac{t_{3}^{(n)}-t_{1}^{(n)}}{t^{(n+1)}_1-t^{(n)}_1}. 
\end{array}
 \quad \mbox{mod~1},  
\label{phaselageq}
\end{equation}
 
 The sequence of phase lags $\left \{ \Delta\phi_{12}^{(n)}, \, \Delta\phi_{13}^{(n)} \right \} $, defined for values between $0$ and $1$, gives the forward phase trajectory on a 2D torus (Fig.~2B). The specific phase-lag values $0$ (or $1$) and $0.5$ represent in-phase and anti-phase relationships, respectively, with the reference cell~1. We examine the $(\Delta\phi_{12},\,\Delta\phi_{13})$-phase space of the 2D Poincar\'{e} return maps (such as one shown in Fig.~3A) of the 3-cell networks by initiating multiple trajectories with a dense distribution of initial phase-lags ($50 \times 50$ grid), and by following their progressions over large numbers of cycles. On long runs these trajectories  can eventually converge to some attractors, one or several. Such an attractor can be  a fixed point (FP) with constant values $\Delta\phi^*_{12}$ and $\Delta\phi^*_{13}$ in (\ref{phaselageq})), which correspond to a stable rhythmic pattern with phase-lags locked (Fig.~2A). All phase trajectories converging to the same fixed point are marked by the same color to reveal the attraction basins of the corresponding rhythms. This reduces the analysis of rhythmic activity generated by a 3-cell network to the examination of the corresponding 2D Poincar\'{e} map for the phase-legs. For example, the map shown Fig.~3A. reveals the existence of penta-stability in the symmetric circuit generating three pacemakers (blue, green and red) and two, clockwise and counter-clockwise, traveling waves (Fig.~3B).  These three PM rhythms  correspond to the blue, green and red FPs located at $(0.5, 0.5)$, $(0.5, 0)$ and $(0, 0.5)$, respectively, while two traveling wave pattern are associated with stable FPs located at  $(1/3, 2/3)$ and $(2/3, 1/3)$, respectively, in the 2D return map.  
 \begin{figure}[t]
\includegraphics[width=0.99\columnwidth]{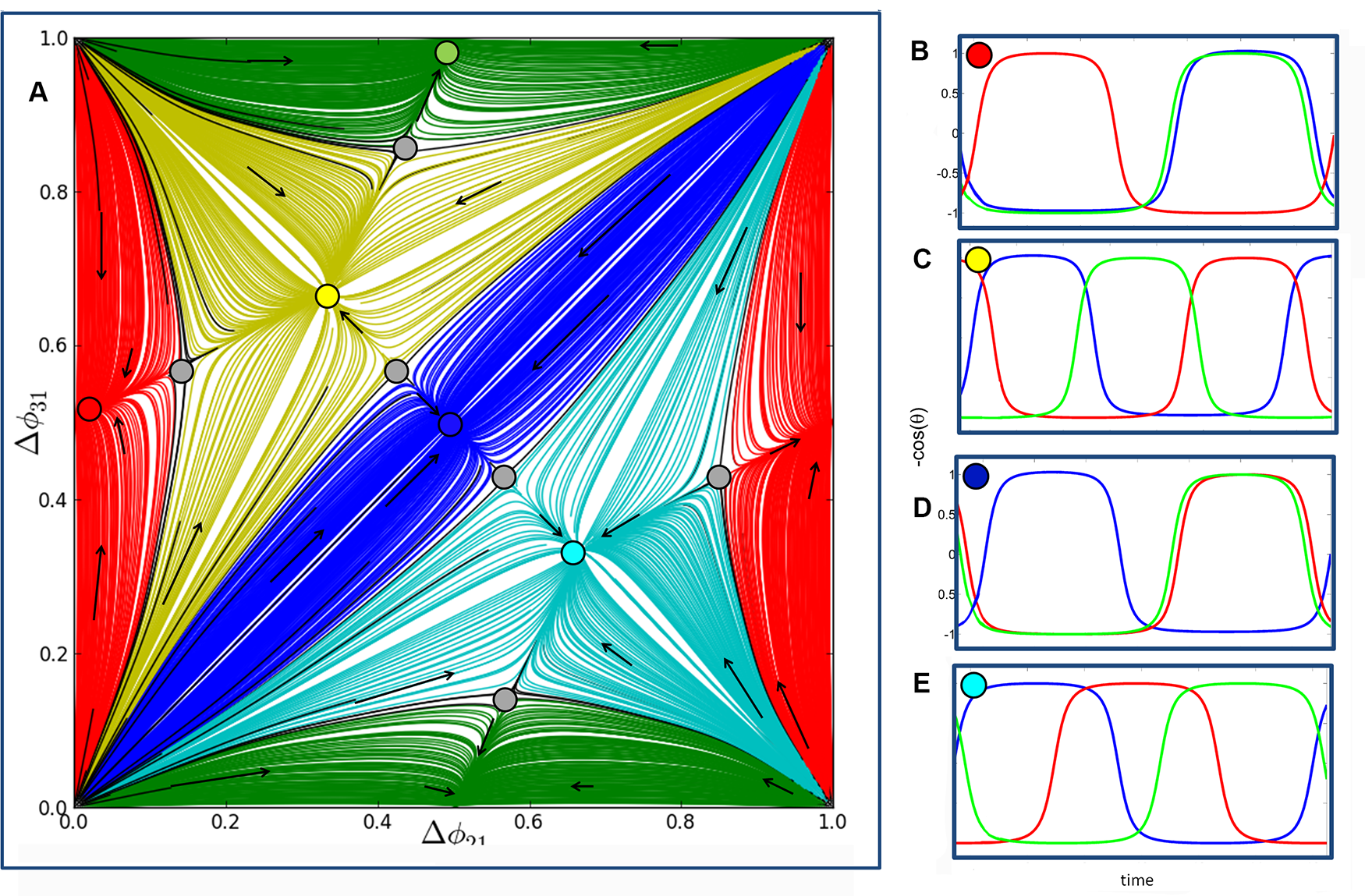}
\caption{Multistable outputs of the 3-cell homogeneous network with six equal synaptic connections ($\beta= 0.003$). (A) The Poincar\'e return map for the $\left (\Delta \phi_{21}, \Delta \phi_{31} \right)$-phase lags with five stable fixed points representing robust three pacemaker (PM) patterns: red at $\left(0,\frac{1}{2}\right)$, green at $\left(\frac{1}{2},0\right)$ and blue at $\left(\frac{1}{2},\frac{1}{2}\right)$, and two traveling wave (TW) rhythmic patterns: yellow clockwise at $\left(\frac{1}{3},\frac{2}{3}\right)$ and teal counter-clockwise at $\left(\frac{2}{3},\frac{1}{3}\right)$. The color-coded attraction basins of these five FPs are determined by positions of stable sets (separatrices) of six saddles (gray dots). The origin is a repelling FP of the map with the even number -- total eight  of hyperbolic FPs in the map. Panels B-E depict the traces with phases locked to the specific values (indicated by color-coded dots at top-left corners), corresponding to the selected FPs.}
\end{figure}  
 Other type of attractors can be a stable invariant curve corresponding to rhythmic pattern wit 
 (a)periodically varying phase-lags. Such a curve can be a circle on and wrap around the 2D torus (see Figs.~2A and 3A).
 If the map has a single attractor, then the corresponding network  is mono-stable, otherwise it is a 
 multifunctional or multistable network capable of producing several rhythmic outcomes robustly. 
 The 2D return map: $M_n  \to M_{n+1}$, for the phase-lags can be represented as follows:
\begin{equation}
\begin{array}{rcl}
\Delta\phi^{(n+1)}_{21}&=&\Delta\phi^{(n)}_{21}+\mu_1 f_1 \left (\Delta\phi^{(n)}_{21}, \Delta\phi^{(n)}_{31} \right ), \\
\Delta\phi^{(n+1)}_{31}&=&\Delta\phi^{(n)}_{31}+\mu_2 f_2 \left (\Delta \phi^{(n)}_{21}, \Delta\phi^{(n)}_{31} \right )
\end{array}
\end{equation}
with small $\mu_i$ being associated with weak coupling; $f_i$ are some undetermined coupling functions such that their zeros: $f_1=f_2=0$ correspond to fixed points: $\Delta \phi^{*}_{j1}=\Delta \phi^{(n+1)}_{j1}=\Delta \phi^{(n)}_{j1}$ of the map. These functions, similar to phase-resetting curves, can be numerically evaluated from the simulated data on all trajectories $\left \{ \Delta\phi^{(n)} _{21}, \Delta \phi^{(n)} _{31} \right \}$ (see Fig.~4C). By treating $f_i$ as partials $\partial F / \partial \phi_{ij}$,  one may try to  restore a ``phase potential''  --  some surface $F \left (\phi_{21}\, , \phi_{31} \right)=C$  (see Fig.~4). The shape of such a surface defines the location of critical points associated with FPs -- attractors, repellers and saddles of the map. With this approach one can try to predict bifurcations due to landscape transformations and therefore to interpret possible  dynamics of the network as a whole. Figure~4A and B are meant to give an idea how the potential surface may look like in the case of the 3-cell  circuit with only two stable traveling wave patterns and in the case of three co-existing pacemakers only, respectively.   
\begin{figure}[t!]
\includegraphics[width=1\columnwidth]{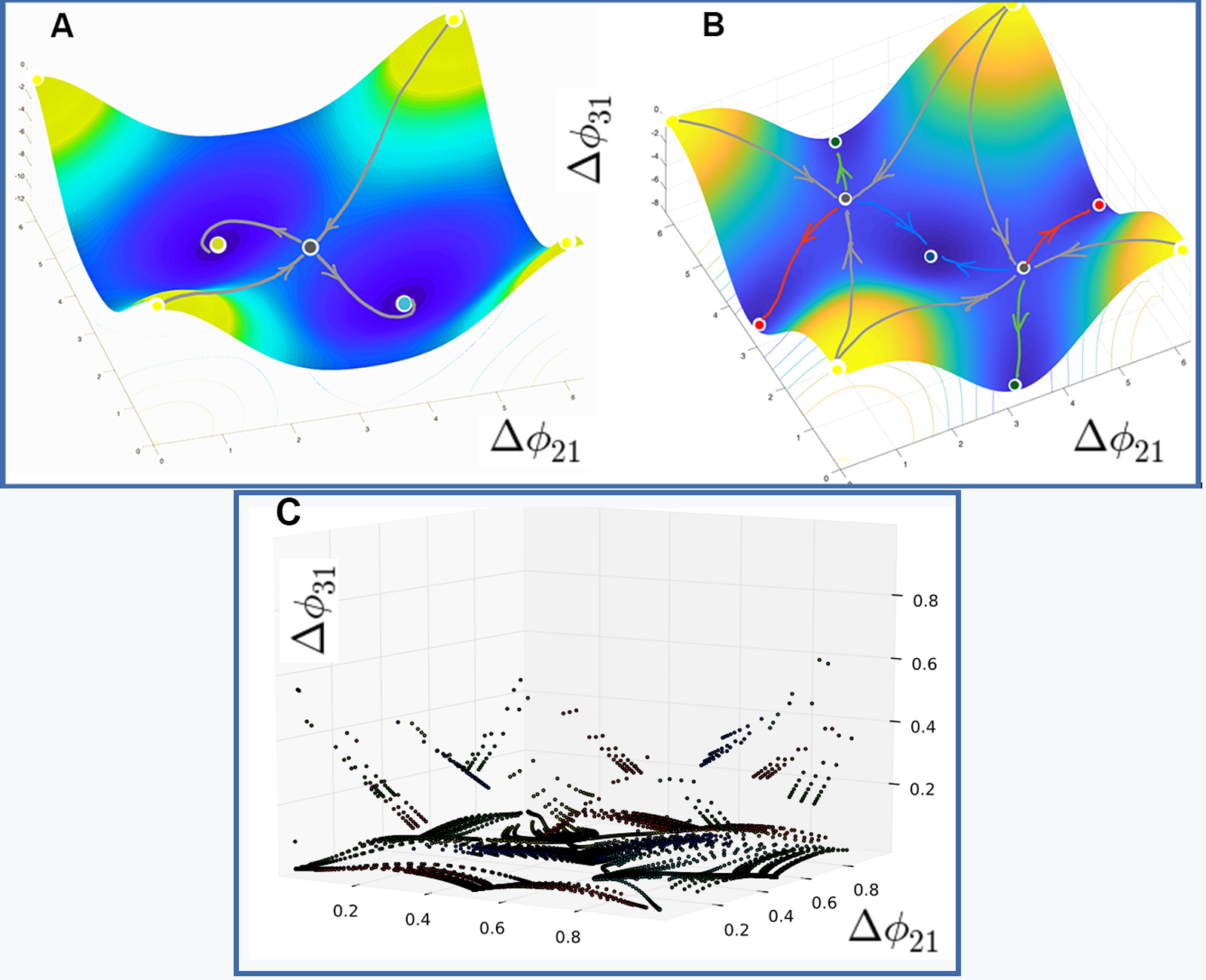}
\caption{Critical points of the sketched ``pseudo-potentials'' with periodic boundary conditions  reveal  the location of potential dwells -- attractors, as well saddles (including one with six separatrices in (B)) and repellers in the ($\phi_{21},\, \phi_{31}$)-phase surface. These configurations correspond the network with only two traveling waves and with only 3 pacemakers. (C) A computational reconstruction of a pseudo-potential/coupling function corresponding to the return map in Fig.~3A. }
\end{figure}

\section{Minimalistic 2$\theta$-burster}
\begin{figure}[ht!]
\includegraphics[width=0.99\columnwidth]{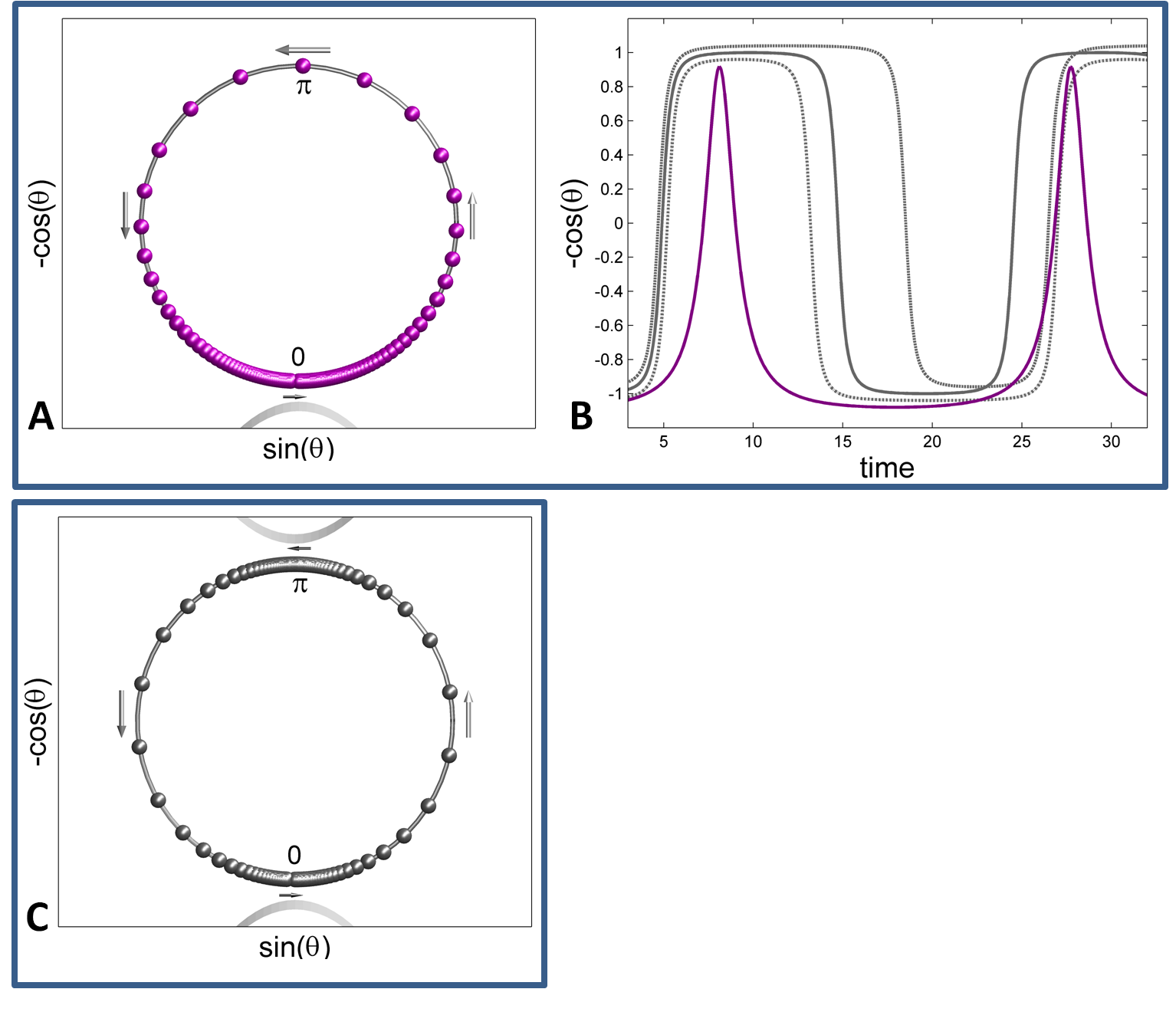}
\caption{Comparison of the oscillatory dynamics generated by the spiking $\theta$-neuron and the $2\theta$-burster. Panels A and C present snapshots of typical trajectories generated by both models on a unit circle ${\mathbb S}^1$ (parametrized using Cartesian coordinates: $x(t)=\sin(\theta(t))$ and $y(t)=-\cos \theta(t)$) with the origin $0$ at 6pm. (A)  Clustering of purple spheres near the origin is due to a bottleneck post-effect caused by a saddle-node bifurcation (SNIC) in the $\theta$ model, while the $2\theta$-burster in (C) features two such bottleneck post-effects due to two heteroclinic saddle-node connections causing the stagnation of gray spheres near the top, ``on'' state  and the inactive ``off''  state of the $2\theta$-burster and fast transitions in between. 
 (B)  Spiking trace (purple) of the $\theta$-neuron, being overlapped with 2-plateau traces of the $2\theta$-neuron with three values of the duty cycles $\simeq$ 50\%,  30\% and 
 70\% (solid, short- and long-dashed gray curves, resp.) }
\end{figure}

The concept of the  $2\theta$-burster is inspired by the dynamics of endogenous bursters with two
characteristic slow phases: depolarized tonic-spiking and quiescent, like one shown in Fig.~1. These phases are often referred to as ``on'' or active and ``off'' or inactive depending on whether the membrane voltage is above or below the synaptic threshold. During the active phase the pre-synaptic cell releases neurotransmitters to inhibit or excite other cells on the network, while during the inactive phase, the cell does not ``communicate'' to anyone. This is a feature of chimerical synapses unlike the electric synapses that let cells interact all the times regardless of the voltage values. The predecessor of the 2$\theta$-burster is the so-called ``spiking'' $\theta$-neuron \cite{theta}.  Mathematically, it is a normal form for the plain saddle-node bifurcation on a circle through which two equilibrium state, stable and repelling, merge and disappear. After the phase point keeps traverse the circle. That is why this bifurcation is referred to as a  homoclinic Saddle-Node bifurcation on an Invariant Circle, or SNIC for short. The $\theta$-neuron capitalizes on the pivotal property of the saddle-node bifurcation -- the phantom bottle-neck effect that underlies slow and fast time scales in the dynamics of such systems, see Fig.~5A.  Recall that a similar saddle-node bifurcation, which controls the tonic-spiking phase and the number of spikes in bursters, is associated with the blue-sky catastrophe \cite{Shilnikov2005a,blue_schol,Shilnikov2012,Showcase2014}.     

\begin{figure}[t]
\includegraphics[width=0.99\columnwidth]{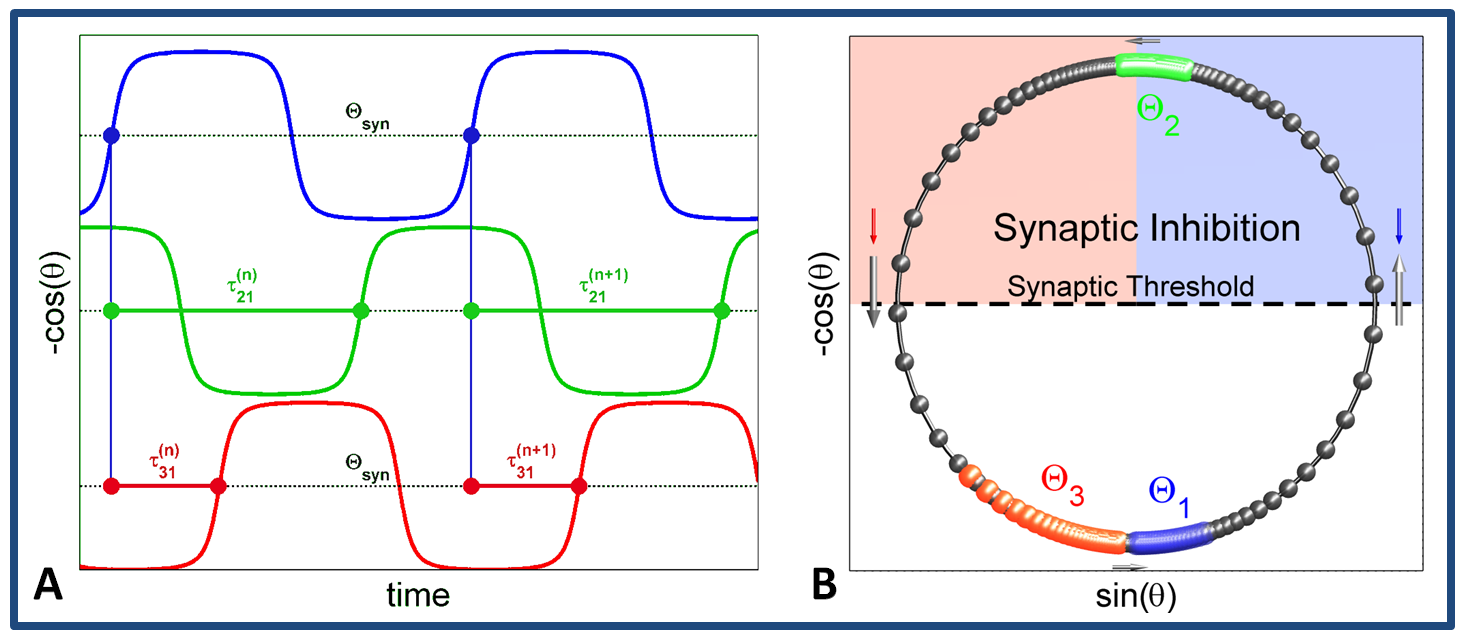}
\caption{(A) Sampling the moments in phase traces,  $y_i (t)=-\cos(\theta_i(t))$, plotted against time, when they reach a synaptic threshold $\theta_{\rm syn}=0$, defines a sequence of the phase lags $\left (\tau_{21}^{(n)}, \tau_{31}^{(n)} \right )$ between upstrokes in the reference, blue neuron and other $2\theta$-neurons coupled in the 3-cell network.
 (B) Parametric representation of the 1D phase space of coupled $2\theta$-bursters traversing counter-clockwise (long gray arrows indicating rapid transition between on-off states) on a unit circle $\mathbb{S}^1$. Small-downward blue and red arrows illustrating the inhibition perturbations from the active green cell above the synaptic threshold  that delays the forthcoming upstroke of the blue cell, and speeds up the red cell toward the inactive phase.}
\end{figure}

The key feature of the $2\theta$-neuron given by 
\begin{equation}
\label{eq:2thetaneuron}
\theta^\prime=\omega-\cos{2\theta}+ \alpha\cos{\theta}, \quad \mbox{mod 1}
\end{equation}
is the occurrence of two saddle-node bifurcations, that introduce two slow phases into its dynamics, with two fast transition in between, see Fig.~5B. Similar to endogenous bursters with two slow transient states -- the active tonic-spiking and the quiescent phases that can be controlled independently, we can manage the durations of the two analogous states in the $2\theta$-neuron: ``on'' at $\pi$  and ``off'' at $0$, using the same bottleneck post-effects of the two saddle-node bifurcations. This allows us to regulate its duty cycle, which is the fraction of the active-state duration compared to the burst period, see Fig.~5B. As seen from Fig.~5, the $\theta$-model was meant to replicate phenomenologically fast spiking cells, while the $2\theta$-neuron can be treated as a ``spike-less'' burster. Below, we demonstrate that the network dynamics produced by a 3-cell motif composed of inhibitory $2\theta$-bursters, preserve all the key features seen in a motif composed of the three Hodgkin-Huxley-type bursters (see Fig.~1).\\         

First, let us observe from Eq.~3 that the phase dynamics of an individual $2\theta$-burster is mainly governed by the terms  $\omega-\cos2\theta$. As long as the frequency $0<\omega \le 1$, there are two pair of stable and unstable equilibria: one pair is at the bottom around $\theta \simeq 0$, while the other is at the top near $\theta \simeq \pi$. The stable states are associated with the hyperpolarized and the depolarized quiescent states of the neuron. When $\omega > 1$, the $2\theta$-burster becomes oscillatory through two simultaneous (provided $\alpha=0$) saddle-node bifurcations (SNIC) on a unit circle ${\mathbb  S}^1$, where $\theta$ is defined on modulo 1. Moreover, as longer as  $\omega=1+\Delta \omega$, where $0<\Delta \ll 1$, this new burster possesses two slow phases: the active ``on'' state near $\theta=\pi$, and the inactive ``off'' state near $0$ on ${\mathbb S}^1$. These slow stated are alternated with fast counter-clockwise transitions, which  are referred to as an upstroke and a downstroke, respectively. For greater values of $\omega$, the active and inactive phases are defined bore broadly:  $\pi/2< \theta \leq 3\pi/2$ and $3\pi/2 < \theta \leq \pi/2$, respectively. This is convenient as the inactive phase remains below the synaptic threshold, which is set at $\theta_{th}=\pi/2$ so that $\cos\theta_{th}=0$ for sake of simplicity, thus equally dividing the unit circle (see Fig.~6A). The duty cycle of the $2\theta$-burster is controlled by the term $\alpha\cos{\theta}$, provided that it remains oscillatory as long as $\omega-|\alpha|>1$. Note that when $\alpha=0$, the duty cycle is 50\% and the burster oscillations have even plagues (see Fig.~5B). The active or inactive phases can be extended or shortened, respectively, with $\alpha<0$, making the duty cycle greater, or vice versa -- the duty cycle of individual burster is decreased with $\alpha>0$.

\section{3 equations for 3-cell network}
A 3-cell circuit of the $2\theta$-bursters coupled with chemical synapses is given by the following system:
\onecolumngrid
\begin{equation}
\label{eq:2thetaneuronfull}
\begin{cases}
\theta'_1=\omega-\cos{2\theta_1}+ \alpha\cos{\theta_1} -  \displaystyle \textcolor{blue} { \left [\frac{\beta_{21}}{1+e^{k\cos{\theta_2}}}+\frac{\beta_{31}}{1+e^{k\cos{\theta_3}}} \right ]} \cdot \textcolor{brown}{ \left [1-\frac{2}{1+e^{k\sin{\theta_1}}} \right ]},\\~\\
\theta'_2=\omega-\cos{2\theta_2}+ \alpha\cos{\theta_2} - \displaystyle \textcolor{blue} { \left [\frac{\beta_{12}}{1+e^{k\cos{\theta_1}}}+\frac{\beta_{32}}{1+e^{k\cos{\theta_3}}} \right ]} \cdot \textcolor{brown}{ \left [1-\frac{2}{1+e^{k\sin{\theta_2}}} \right ]},\\~\\
\theta'_3=\omega-\cos{2\theta_3}+ \alpha\cos{\theta_3} -  \displaystyle \textcolor{blue} { \left [\frac{\beta_{13}}{1+e^{k\cos{\theta_1}}}+\frac{\beta_{23}}{1+e^{k\cos{\theta_2}}} \right ]} \cdot \textcolor{brown}{ \left [1-\frac{2}{1+e^{k\sin{\theta_3}}} \right ]},
\end{cases}
\quad \mbox{mod 1.}
\end{equation}
\twocolumngrid
The $2\theta$-burster are coupled in the network using the fast inhibitory synapses driven by the fast-threshold modulation \cite{FTM}. It is due to the  positive ``sigmoidal'' term  $ \displaystyle \textcolor{blue} { \left[ \frac{1}{1+e^{k\cos{\theta_i}}} \right ] }$ that, rapidly ((here $k=10$)  varying between $0$ and $1$, triggers an influx of inhibition flowing from the  pre-synaptic neuron into the post-synaptic neuron, as soon as the former enters the active on-phase above the synaptic threshold $\cos \theta_{\rm th} =0$, i.e.,  $\pi/2<\theta_i <3\pi/2$. Note that the negative sign of this term makes the synapse inhibitory; replacing it with ``+''  makes the synapse excitatory because it would increase the rate of $\theta'$ during the upstroke, contrarily to slowing the upstroke down as in the inhibitory case. The strength of the coupling is determined by the maximal conductance values $\beta_{ij}$.  

The last term  $\textcolor{brown}{ \left [1-\frac{2}{1+e^{k\sin{\theta}}} \right ]}$, breaking the symmetry, 
converts the synaptic input into qualitative inhibition. Namely, its sign is switched from + to - upon crossing the values $\theta = 0$  and $\theta = \pi$. During the fast upstroke, when $0 <\theta <\pi$, the this term is positive, thereby ensuring that inhibition does slow down or delay the transition into bursting. When $\pi <\theta_i <2\pi$ during the fast downstroke, this terms $\textcolor{brown}{ \left [1-\frac{2}{1+e^{k\sin{\theta}}} \right ]}<0$ to unsure that the inhibition speeds up the transition from the active (tonic-spiking) phase bursting into the inactive (quiescence) phase faster. This is phenomenologically  consistent with neurophysiological recordings  as inhibition projected onto the post-synaptic burster typically shortens the burst duration and extends the interburst intervals. One can alternatively replace this term with $\textcolor{brown} { \left [1-\frac{1}{1+e^{k\sin{\theta}}} \right ]}$ that breaks the symmetry was well as it acts during the upstroke only. 

The electrical coupling or the gap junction between the neurons is handled by the another term $ -C_{\rm elec}\sin(\theta_{\rm pre} - \theta_{\rm post}) $. It slows down the rate $\theta'_{post}$ when $\theta_{post} > \theta_{pre}$ and speeds it up if $\theta_{post} < \theta_{pre}$. The conductivity coefficient $C_{\rm elec}$ is to be set around two orders of magnitude smaller than $\beta$-values to maintain a balanced effect in the network. When  $C_{\rm elec}$ and $\beta$ are of the same magnitude, the dynamics of network are solely dictated by the electrical coupling with the inhibitory synapses insignificantly affecting it.

\section{Poincar\'e return maps for the phase-lags. Results}

\begin{figure*}[t]
\includegraphics[width=1.6\columnwidth]{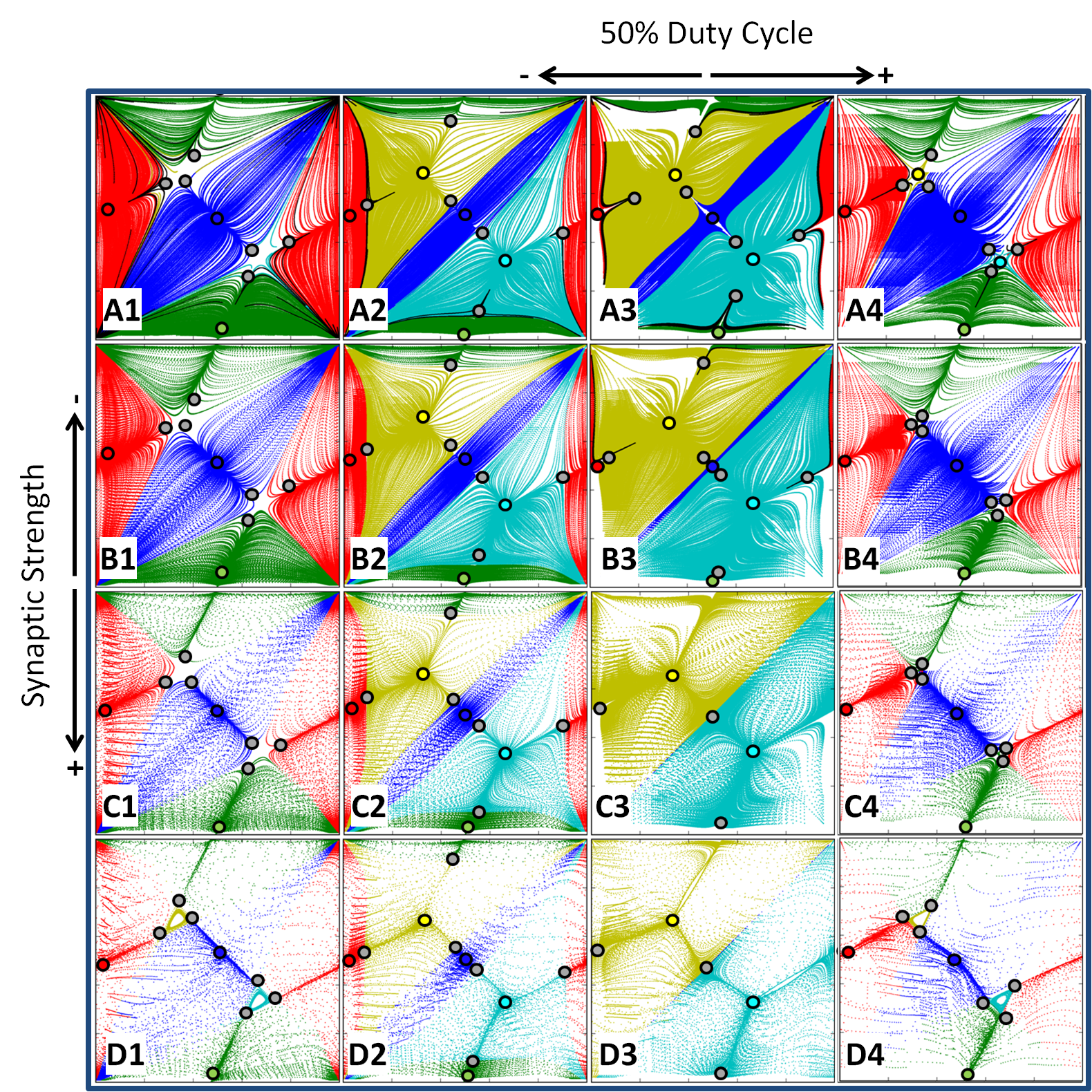}
\caption{Bifurcations of FPs in the $(\Delta \phi_{21}, \Delta \phi_{31})$-return map for the symmetric motif as the coupling $\beta$ -parameter and the duty cycle (via variations of $\alpha$) are changed; parameters: $\beta$-values are [0.001\,,0.003,\,0.01,\,0.03] from top to bottom labeled A to D, resp., while  $\alpha$-values are [-0.11,-0.05,0.0,0.11] from left to right labeled, 1 through 4, respectively, with 50\% DC at $\alpha=0.0$ in column 3. With larger $\beta$-values, the rate of convergence to the FPs increases. The TW-rhythms dominate the network dynamics when the DC is about 50\%, as seen in the middle columns. The PM-rhythms become dominant at small and large DC-values, as depicted in the outer panels.}
\end{figure*}

Figure~6 shows the snapshots of the phase progressions of the three coupled $2\theta$-bursters on the unit circle ${\mathbb S}^1$ and depicts how phase-lags between the are introduced (here, we refer to cell 1 (blue) as the reference one). One can see from this figure that the active green neuron~2 in the active phase close to $\theta=\pi$, above the synaptic threshold, inhibits and pushes the other two closer to each other, near the bottom quiescent state at $\theta=0$, by accelerating the red burster~3 on the downstroke, and by slowing down the blue burster~1 on the upstroke.

Following the same approach used in the weakly coupled HH-type models above, we first create a uniform distribution of initial phases on ${\mathbb S}^1$, and therefore the phase-lags between the three $2\theta$-bursters. Next we integrate the network~(4) over a large number of cycles, and record burst initiations (see Fig.~5A) to determine the phase-lags between the reference cell~1 and two other cells and to what phase locked states they can converge  with increasing number of the cycles. This approach is illustrated in Fig.~2A for the symmetric 3-cell motif composed of identical $2\theta$-bursters and equal inhibitory synapses. The corresponding 2D Poincar\'e return map, with the co-existing stable fixed points and saddles is shown in Figs.~3. By stitching together the opposite sides of this map, we wrap it around a 2D torus as  shown in Fig.~2B. 

The fixed points and their attraction basins are coded with different colors in the map. For example, the Poincar\'e return map for the $\left (\Delta \phi_{21}, \Delta \phi_{31} \right)$-phase lags represented in Fig.~3A has five stable fixed points representing robust three pacemaker FPs located at: red $\left(0,\frac{1}{2}\right)$, green at $\left(\frac{1}{2},0\right)$ and blue at $\left(\frac{1}{2},\frac{1}{2}\right)$, and two traveling-wave ones: yellow clockwise at $\left(\frac{1}{3},\frac{2}{3}\right)$ and teal counter-clockwise at $\left(\frac{2}{3},\frac{1}{3}\right)$. The borders of the  attraction basins of these five FPs are determined by positions of stable sets (separatrices) of six saddles (gray dots). The origin is a repelling FP of the map. It totals up to eight hyperbolic FPs in the map. 

To conclude this section, we would like to point out another helpful feature of the $2\theta$-neuron paradigm, namely, the ability to find repelling FPs, if any, in the 2D Poincar\'e map, by reversing the direction of integration of the system~(4), i.e., integrating it in the backward time after multiplying its  right-hand sides by -1. Unlike HH-type dissipative systems where the backward integration will make solutions run to infinity, it is not the case for the $2\theta$-bursters, as just reverses the direction and spin trajectories clockwise on ${\mathbb S}^1$.

\subsection{Symmetric Motif}

It will be shown below that the $2\theta$-bursters weakly coupled in the 3-cell networks, symmetric, asymmetric and with mixed synapses, can generate the same stable rhythms as the networks of biologically plausible HH-type models. We will also discuss the bifurcations occurring in the networks and corresponding maps as the synaptic connectivity or intrinsic temporal characteristics of the $2\theta$-bursters are changed. Bifurcations in the system are identified and classified by a change of the stable phase rhythms which can be due to the stability loss of a particular FP, or when it merges with a close saddle so both  disappear through a saddle-node bifurcation. 

\begin{figure}[t]
\includegraphics[width=0.7\columnwidth]{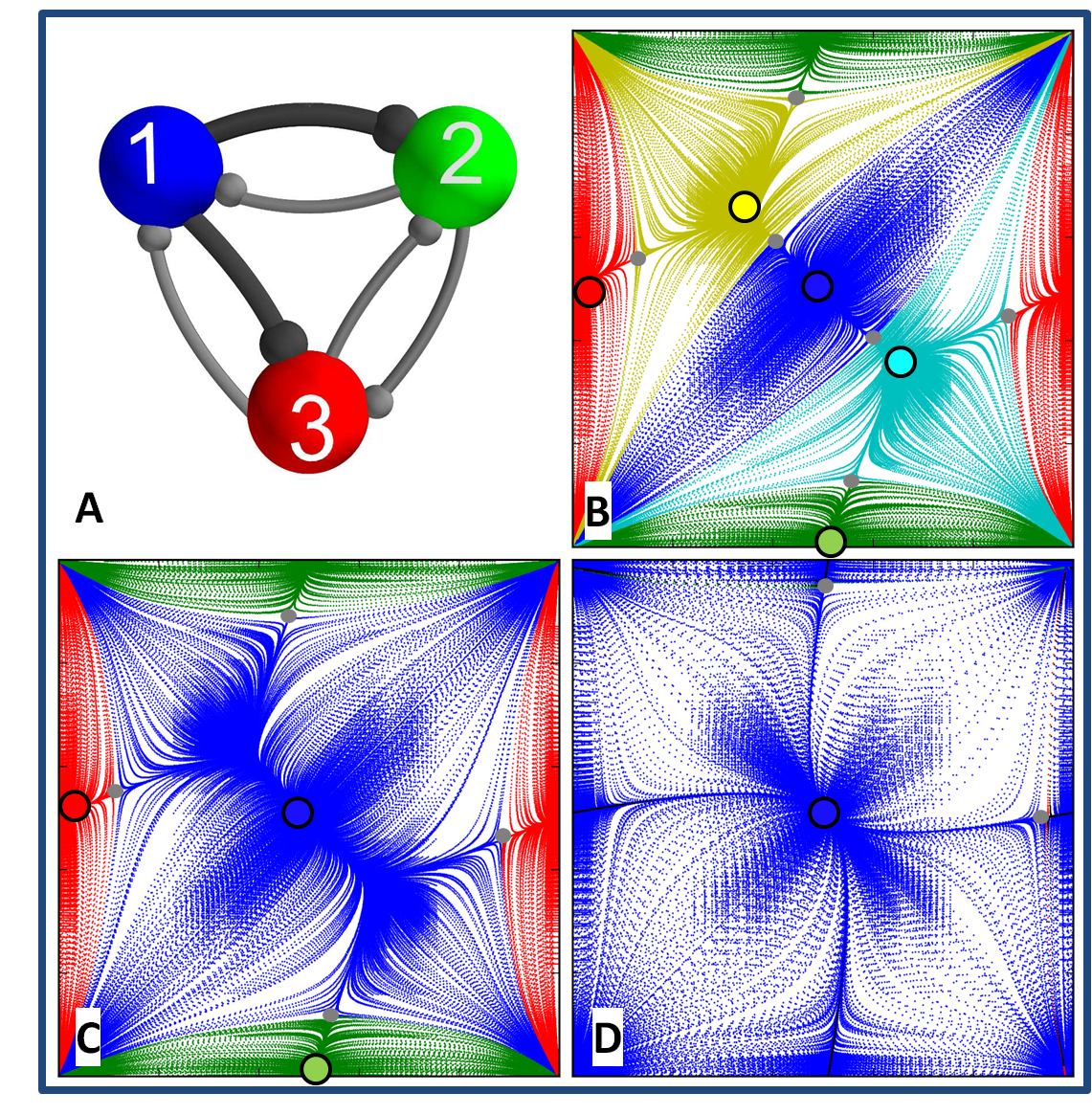}
\caption{(A) ``King of the mountain'' network motif with two synapse strengths, $\beta_{13}$ and $\beta_{12}$, increased (indicated by darker connections), relative to the other synapse strengths. (B) The first of three $\left (\Delta \phi_{21}, \Delta \phi_{31} \right)$ return maps, with $\beta_{13}$ and $\beta_{12}$ synaptic strengths slightly greater than the other $\beta$s, the (blue) attraction area extends so that the two saddles nearest the blue PM at $\left(\frac{1}{2},\frac{1}{2}\right)$, move away from the blue PM, closer towards the yellow and teal TWs at $\left(\frac{1}{3},\frac{2}{3}\right)$ and $\left(\frac{2}{3},\frac{1}{3}\right)$, respectively. (C) With further increase of $\beta_{13}$ and $\beta_{12}$, these saddles and TWs merge with and annihilate each other through saddle-node bifurcations, and the blue PM basin grows. (D) At greater $\beta_{13}$ and $\beta_{12}$ values, the network becomes a winner-take-all, blue PM winning, after the red and green PMs, at $\left(\frac{1}{2},0\right)$ and $\left(0,\frac{1}{2}\right)$, respectively, vanish through subsequent saddle-node bifurcations. The parameters are: $\omega$ = 1.15, $\alpha$ = 0.07, and $\beta$ = 0.003 except $\beta_{13}$ and $\beta_{12}$ = 0.0038, 0.004, 0.015 for panels B-D.}
\end{figure}

Let us first consider a symmetric network with two bifurcation parameters: the coupling strength $\beta=\beta_{ii}$ $(i=1,2,3)$ and the $\alpha$-parameter in Eq.~(3) that controls the duty cycle (DC) of the $2\theta$-bursters. We use five different DC-values as $\alpha$ is varied from -0.11 to 0.11l while  synaptic strength is increased through four steps from $\beta$ = 0.0001 through 0.1. The results are presented in Fig.~7. The Panels A2/3 represent the most balanced, weakly coupled network that can produce all five bursting rhythms with the DC ~50\%.  One can see that increasing the $\beta$-value, the saddles separating 2 TWs and 3 PMs move toward the latter ones, and  over some critical values, 3 pairs: a saddle and the nearest stable PM merger and vanish simultaneously. After that, the symmetric network can produce two only rhythms: counter- and clockwise TWs corresponding to the teal and yellow stable  FPs at $\left(\frac{1}{3},\frac{2}{3}\right)$ and $\left(\frac{2}{3},\frac{1}{3}\right)$, respectively. This would correspond to the case of the ``pseudo-potential'' depicted in Fig.~4A.         
     
The stable PMs are promoted or dominate the dynamics of the symmetric at  the extreme $\alpha$-values corresponding to the bursting rhythms with short or long burst durations.  Once can compare panels, say A1 and D4 reveal that this time, the separating saddles group around the stable TWs to minimize their attraction basins, and hence the likelihood of the occurrence of these rhythms in the network. These case would correspond to the ``pseudo-potential'' depicted in Fig.~4B.          

\subsection{``King of the mountain'' motif}

The first asymmetric case considered is a motif termed the King of the Mountain. In this modeling scenario both outgoing inhibitory synapses from the given cell, here the reference blue burster~1 one, are evenly increased in the strength, see Fig.~8A. Observe that such a configuration breaks down both circular symmetries supporting traveling waves in the network.  Let us start with Fig.~8B: no surprise that with initial increase in $\beta_{1,2/3}$, two saddles shift away from the blue PM at (0.5,0.5)  toward two TWs, then merge with them to disappear pair-wisely. Next, as $\beta_{1,2/3}$ is increased further, two other saddles annihilate the green and red PMs through similar saddle-node bifurcations (Fig.~8C). At the aftermath, the 3-cell network with a single burster generating the repulsive inhibition much stronger than the other two cells becomes a monostable one producing a single pacemaking rhythm with the phase-lags locked at (0.5,0.5).

\subsection{Mono-biased motif}

\begin{figure}[t]
\includegraphics[width=0.99\columnwidth]{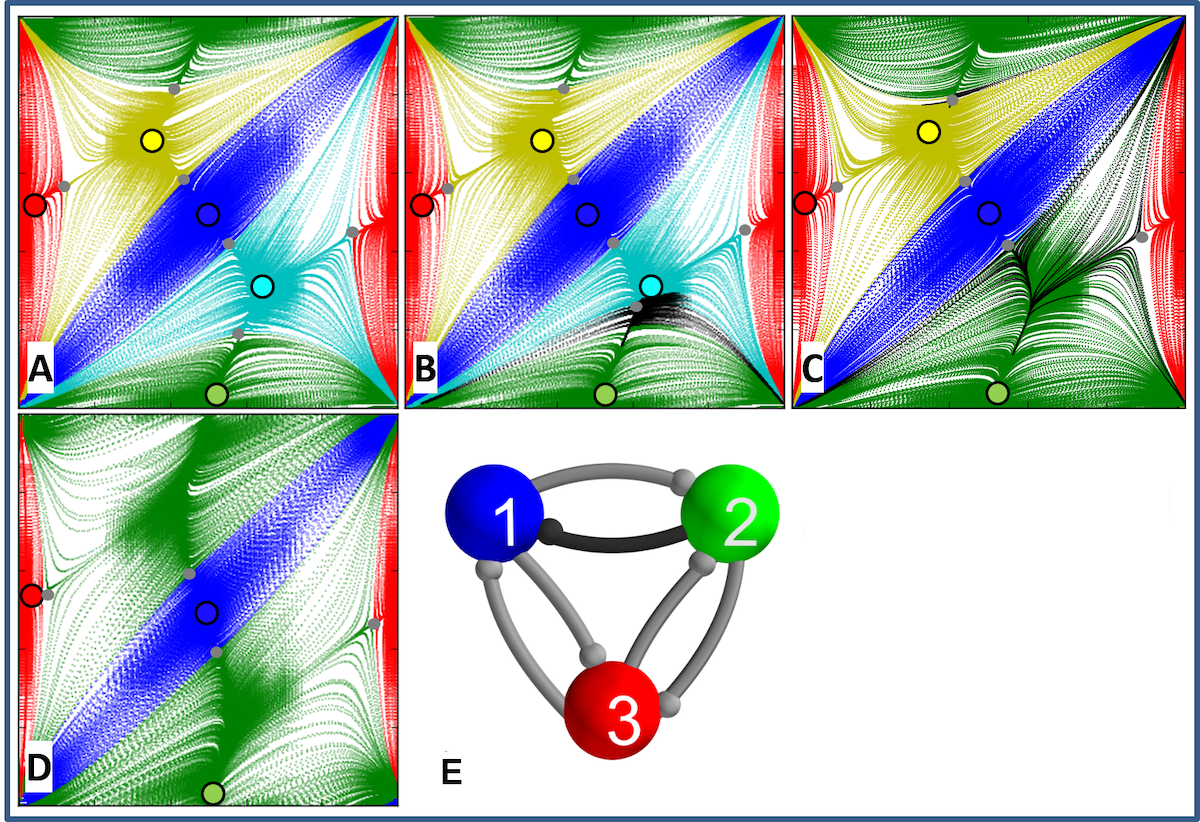}
\caption{Mono-biased network motif (F) with one different synapse due to increasing  $\beta_{21}$. (A) The first of five $\left (\Delta \phi_{21}, \Delta \phi_{31} \right)$ return maps, an increase in $\beta_{21}$ value breaks down a counter-clockwise symmetry so that the attraction basin (teal) of the corresponding  TW at $\left(\frac{2}{3},\frac{1}{3}\right)$ shrinks as a nearby saddle moves closer to it and away from the green PM at $\left(\frac{1}{2},0\right)$ (A and B). (C) With further increase of $\beta_{23}$, the counter-clockwise TW at $\left(\frac{2}{3},\frac{1}{3}\right)$ vanishes through a saddle-node bifurcation after merging with the nearby saddle, followed by another saddle-node bifurcation eliminating the red PM at (0,\,0.5) (D). At greater $\beta_{23}$ values the green PM $\left(\frac{1}{2},0\right)$ encompasses the majority of the network phase space, along with the blue PM at $\left(\frac{1}{2},\frac{1}{2}\right)$ preserving the size of its attraction basin. The parameters are: $\omega$ = 1.15, $\alpha$ = 0.07, and $\beta$'s = 0.003 except $\beta_{21}$ = 0.00042, 0.0045, 0.01, 0.02 for panels A-D.}
\end{figure}

We refer as a mono-biased motif to another asymmetric the network with a single different synapse: in this case the strength $\beta_{21}$ of the outgoing synapse from cell~2 to cell~1 is increased, which violates the circular symmetry supporting the counter-clockwise traveling wave in the network, see Fig.~9F.  So, as $\beta_{21}$  is increased the counter-clockwise stable FP at $\left(\frac{2}{3},\frac{1}{3}\right)$ first disappears through a saddle-node bifurcation, as seen in  Fig.~9A/B. Because this was the saddle between this TW and he green PM, then the attraction basin of the latter increases after the first bifurcation in the sequence. The next saddle-node bifurcation eliminates the red stable FP at (0,\,0.5). The reasoning is the following: for this rhythm to persist the red PM is to evenly inhibit both green and blue PMs. However, a growing inhibition misbalance between them is no longer reciprocal.  As we pointed out earlier, the stronger inhibition from cell~2 shortens the active phase of the blue burster. As so they cannot be longer lined up by the burster~3, which causes the disappearance of this PM-rhythm and the FP itself (Fig.~9C).  Same arguments can be just to justify the the disappearance of the green PM as cell~2 cannot not even inhibit cells 1 and 2 to hold them together as $\beta_{21}$  is increased further (not shown). This is in the his case is in good agreement with the 3-cell networks of the HH-type bursters.

\begin{figure}[t]
\includegraphics[width=0.99\columnwidth]{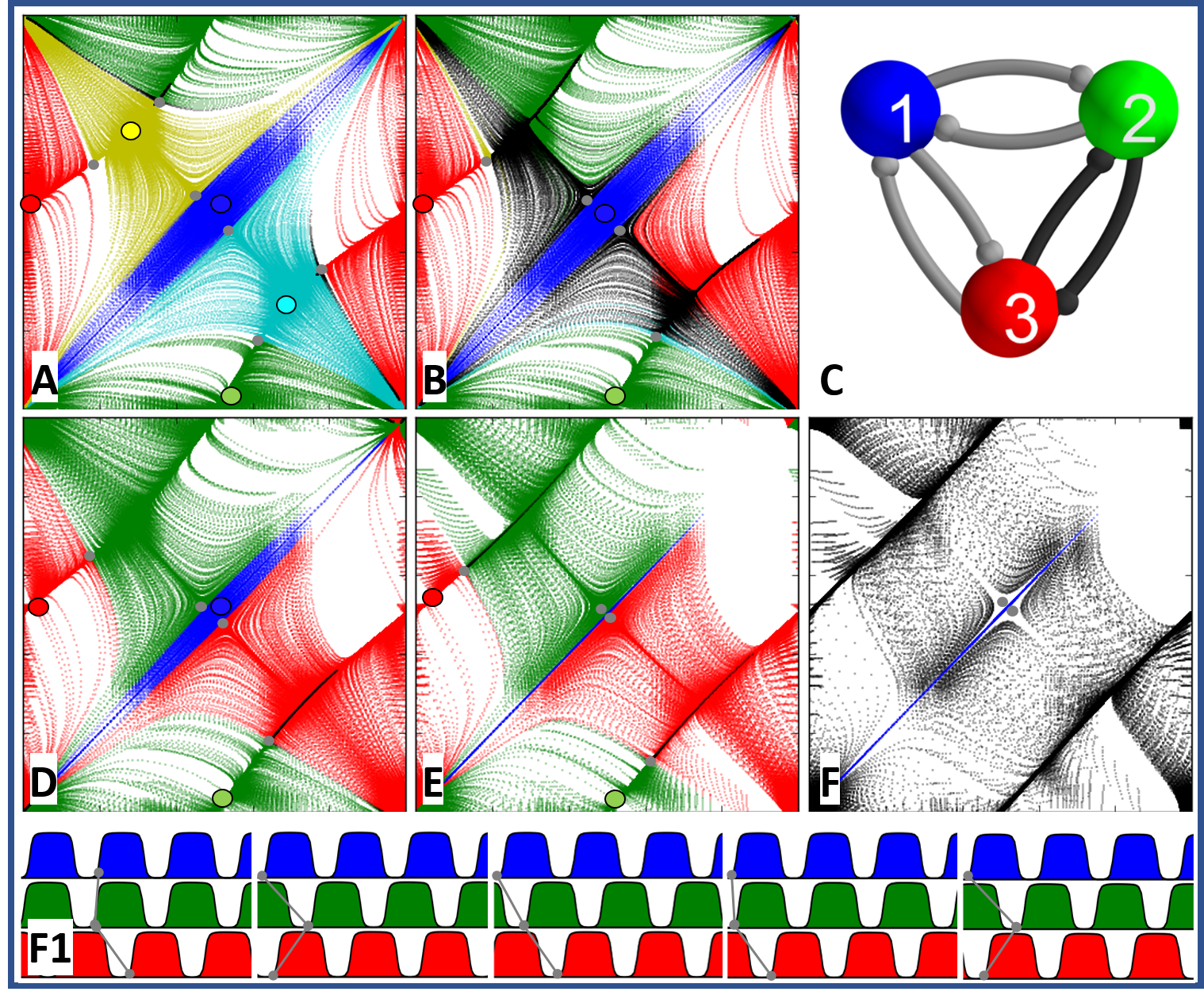}
\caption{(C) ``Pairwise-biased'' network motif with two reciprocal synapse strengths $\beta_{23}$ and $\beta_{32}$, increased.  (A) The first of five $\left (\Delta \phi_{21}, \Delta \phi_{31} \right)$ return maps, with $\beta_{23}$ and $\beta_{32}$ slightly greater than other synaptic connections the network possesses all five attracting FPs. (B) Evenly increasing $\beta_{23}$ and $\beta_{32}$ values breaks down the rotational symmetry of the network so that both TWs at $\left(\frac{1}{3},\frac{2}{3}\right)$ and $\left(\frac{2}{3},\frac{1}{3}\right)$ vanish through saddle-node bifurcations while that the red and green PM basins equally expand and the blue basin shrinks. Here, two areas of the map, due to slow transitions throughout the saddle-node ghosts, are color-coded in black because of uncertainty in ultimate convergence/destination. (D-E) With further increases of $\beta_{23}$, $\beta_{32}$ values,  the blue basin continues to shrink until red and green basins encompass almost all of the areas of the map. One can see from Panel E that that the red and green PMs at $\left(\frac{1}{2},0\right)$ and $\left(0,\frac{1}{2}\right)$ are also about to merge with nearby saddles and disappear through two homoclinic saddle-node bifurcations (SNIC). (F) At greater values of $\beta_{23}$, $\beta_{32}$, the blue PM at $\left(\frac{1}{2},\frac{1}{2}\right)$ has only a very narrow attraction basin, corresponding to the only phase-locked rhythm, co-exists with a dominant {\em phase-slipping} repetitive pattern. The phase slipping (its trace shown in Panel F1) corresponds to a stable invariant curve, passing throughout $\left(\frac{1}{2},0\right)$ and wrapping abound the 2D toroidal phase space to re-emerge near $\left(0,\frac{1}{2}\right)$ and so forth. (F1) Five exemplary episodes of the traces vs. time showing periodically varying phase lags between three cells (slipping).  The parameters are: $\omega$ = 1.15, $\alpha$ = 0.07, and $\beta$ = 0.003, except $\beta_{23}$ and $\beta_{32}$ are 0.005, 0.006, 0.009, 0.035, in panels A, B, D-F and F1 is equal to F.}
\end{figure}

\subsection{Dedicated HCO}

The abbreviation HCP stands for a half-center oscillator, which a pair of neurons coupled reciprocally by inhibitory synapses to produce alternating bursting. Such a dedicated  HCO is formed by cells 2 and 3 with stronger synapses due to $\beta_{23}=\beta_{32}$ in the configuration shown in Fig.~10C. Again with start off with the symmetric case depicted in Fig.~10A. One can observe at once, that having the dedicated HCO should breaks down the circular symmetries of the network. So, the stable TWs become eliminated first as $\beta_{23}=beta_{32}$ starts increasing. As these synapses become stronger the attraction basin of the blue PM at (0.5\,0.5) shrinks substantially, but the FP itself persists. Meanwhile increasing $\beta_{23}=beta_{32}$ further creates the inhibitory misbalance that males the further existence of the green and red PMs impossible due to the factors that we outlines above for the mono-biased motif. Both vanish at the same time due to saddle-node bifurcations. However, at the bifurcation both double FPs are connected by a heteroclinic orbit that transforms into a stable invariant curve wrapping around the phase torus (Figs 10F and 2B). This stable invariant curve is associated with a phase-slipping rhythms that recurrently passes slowly through the ``ghosts'' of all four vanished FPs except for the coexisting blue PM, see the fragments of traces in Fig.~10F.

\subsection{Clockwise-biased motif}

The clockwise-biased motif in this case represents the 3-cell network canter-clockwise  connections stronger than ones in the opposite direction, see Fig.~11E. This configuration does not break circular symmetries of the network but infers that either TW should gain over the opposite one, which should result in that their attraction basins should change correspondingly. Figure~11 presents four transformation stages of  the map as $\beta_{13}$, $\beta_{32}$ and $\beta_{21}$ sequentially  increased. With a small increase, the shape of the map becomes a bit twisted with the three saddles shifting away from the stable PMs toward the teal TW at  $\left ( \frac{2}{3},\frac{1}{3} \right )$. The further increasing brings the saddle close to the latter one thereby shrinking its attraction basin and substantially widening the basin of the clockwise TW at $\left( \frac{1}{3},\frac{2}{3} \right )$. Finally, as some bifurcation threshold is reached, the saddles collapse at the stable FP that becomes a complex saddle with three outgoing and three incoming separatrices. This means that the counter-clockwise TW becomes an unstable rhythm in such biased 3-cell motif that is fully dominated by the  clockwise TW rhythm. 
\begin{figure}
\includegraphics[width=0.99\columnwidth]{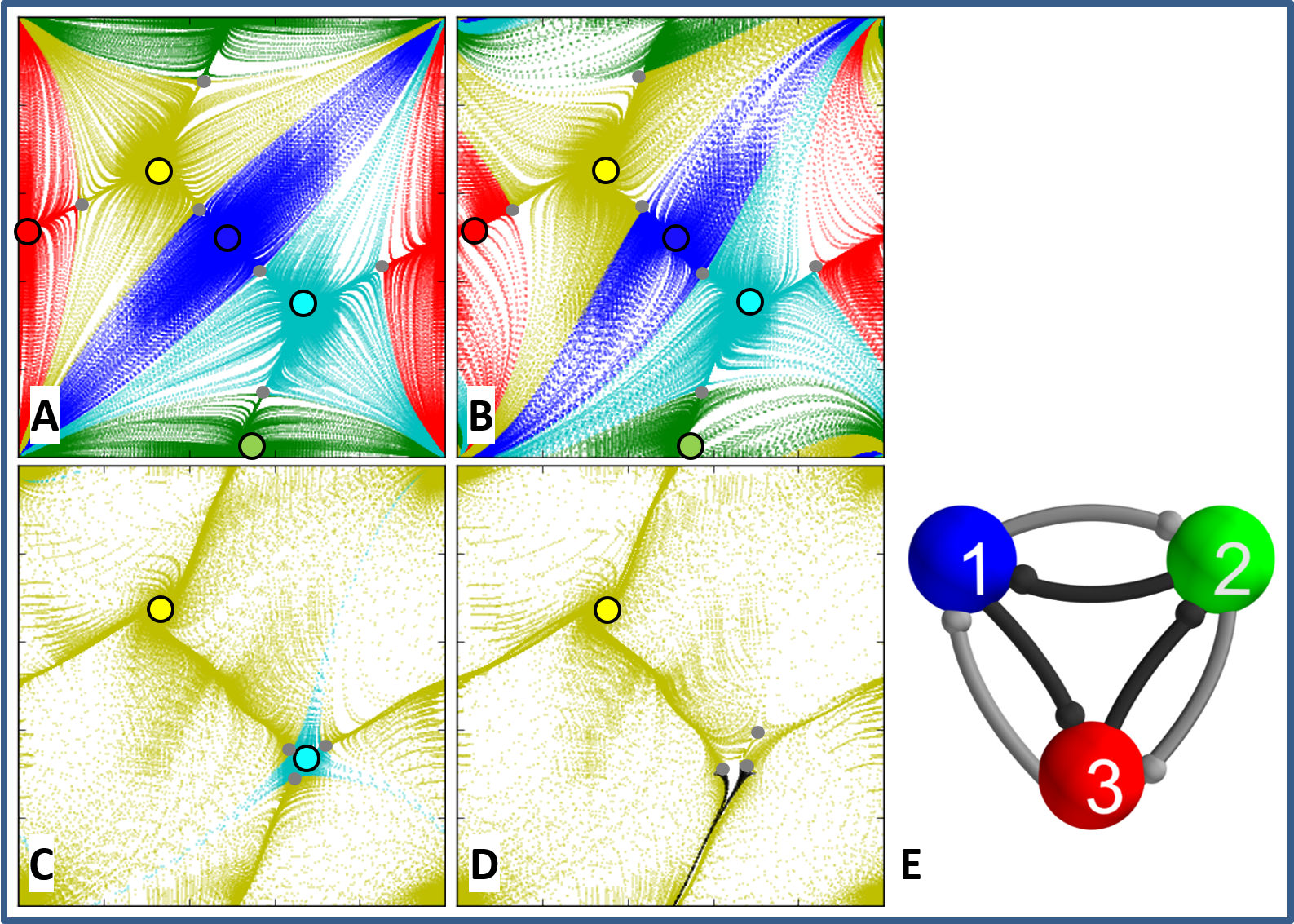}
\caption{(E) Clockwise-biased motif with three synaptic strengths, $\beta_{13}$, $\beta_{32}$ and $\beta_{21}$ sequentially  increased. (A) As all three counter-clockwise synapses are slightly strengthen, saddles shift away from the three stable PMs, blue at $\left( \frac{1}{2}, \frac{1}{2} \right )$, green $\left ( \frac{1}{2} , 0 \right )$ and red $\left ( 0, \frac{1}{2} \right )$,  towards the teal clockwise TW at $\left ( \frac{2}{3},\frac{1}{3} \right )$ (B) thus shrinking its basin and widening the attraction basin of the dominant counter-clockwise TW (yellow) at $\left( \frac{1}{3},\frac{2}{3} \right )$ (C).  (D) With the stronger synaptic values,  the three saddles collapse into the CC TW, which becomes a complex saddle with three incoming and three outgoing separatrices. The parameters are $\omega$ = 1.15, $\alpha$ = 0.07, $\beta$ = 0.003 except $\beta_{12}$, $\beta_{23}$ and $\beta_{31}$ = 0.0033, 0.025, 0.035, 0.055 for panels A-D.}
\end{figure}

\subsection{Gap junction}

In out last example we consider the symmetric motif with a gap junction or an electric synapses added between cells 1 and 2 as shown in Fig.~12C. Recall that a gap junction is bi-directional unlike uni-directional chemical synapses with synaptic thresholds. Recall that it is modeled by this  term $ -C_{\rm elec}\sin(\theta_{\rm pre} - \theta_{\rm post})$ that slows down the rate $\theta'_{post}$ when $\theta_{post} > \theta_{pre}$ and speeds it up if $\theta_{post} < \theta_{pre}$. Due to this property, the electrical like excitatory synapse promote synchrony between such coupled oscillatory cells, which in our case between cells 1 and 2.        

 Observe that introducing an electrical synapse between only two of the cells of the motif ruins both circular symmetries in the system.  This is documented in Fig.~12A/B depicting the maps for the networks with $C_{elec}$ being increased from zero to $0.0003$. Once can see that both TWs were first to vanish from the repertoire of the network. Further increase of $C_{\rm elec}$ makes the stable green and blue stable PMs disintegrate as both cells become synchronous to  burst in alternation with the red cell~3. This completes the consideration of the mono-stable network with a relatively strong gap junction between cells 1 and 2 that can only produce the only one pacemaker rhythm.

\begin{figure}
\includegraphics[width=0.7\columnwidth]{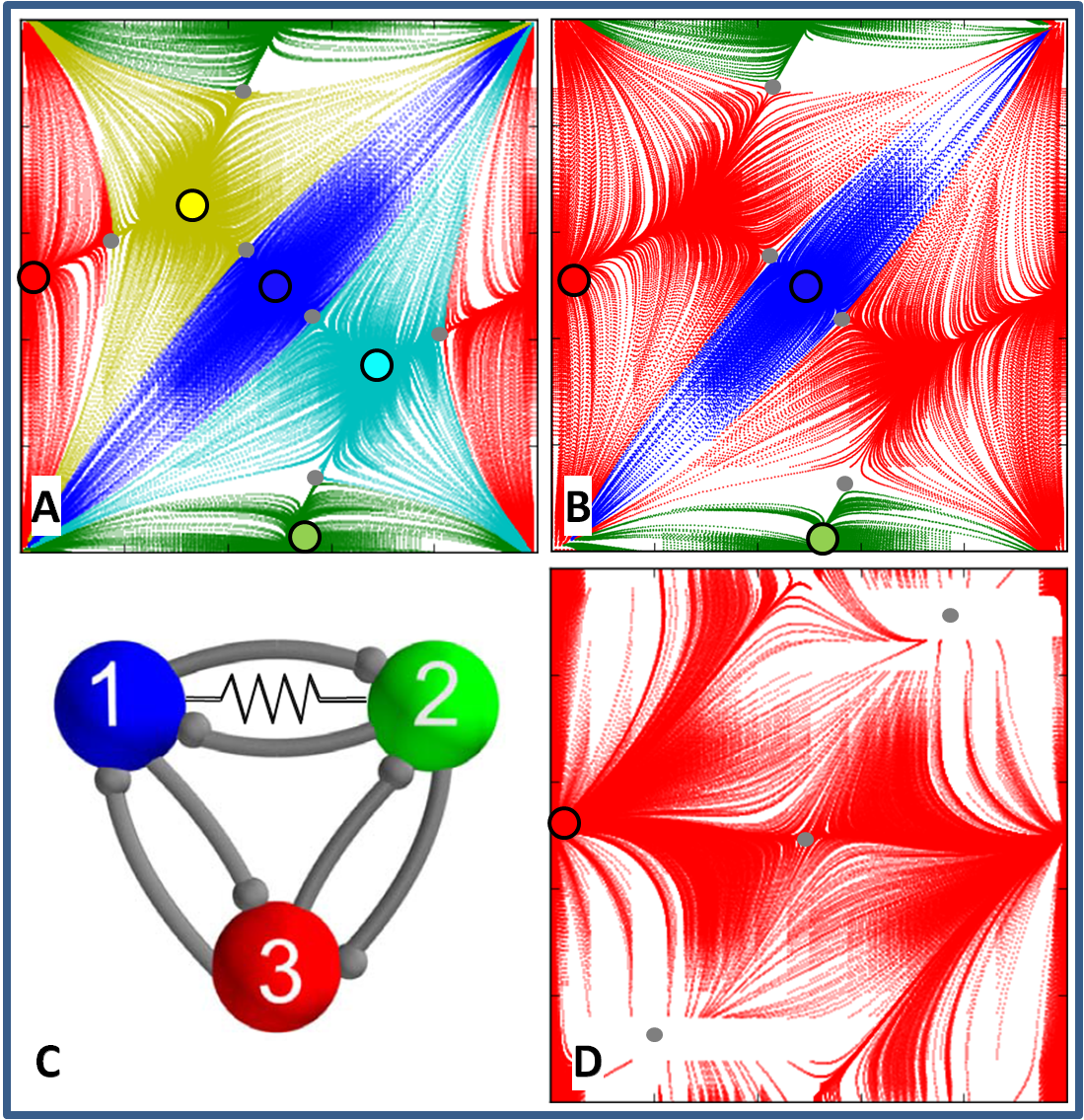}
\caption{Gap junction in the symmetric 3-cell network (C) is represented by a resistor symbol placed between cells 1 and 2. (A) At $C_{elec}=0.00015$ the network yet generates five phase-locked rhythmic rhythms with comparably sized basins of attraction. (B) Increased $C_{elec}$  breaks the circular symmetries of the network that makes both TWs at $\left( \frac{1}{3}, \frac{2}{3} \right )$ and $\left( \frac{2}{3}, \frac{1}{3} \right )$ vanish through saddle-node bifurcations while the basin of the red PM at $\left( 0, \frac{1}{2} \right )$ widens. (D) With an even greater electrical coupling the red PM becomes the winner-takes-all after the electrical connection ensures the in-phase synchrony between cells 1 and 2 (C) that eliminates the blue and green PMs in the map after subsequent saddle-node bifurcation. The parameters are: $\omega$ = 1.15, $\alpha$ = 0.07, $\beta$ = 0.003, and  $C_{elec}=0.00015$, 0.0003, 0.0015 for panels A, B, and D.}
\end{figure}

\section{Discussion}

The goal of this paper is to demonstrate the simplicity and usability of the 2$\theta$-bursters to construct multistable, polyrhythmic neural networks that have the same dynamical and bifurcation properties as ones composed of biologically plausible models of Hodgkin-Huxley type bursters and synapses.  Our de-facto approach is based on the computational reduction to the visually evident Poincar\'e return maps for phase lags derived from multiple voltage traces. These maps serve as a detailed blueprint containing all necessary information about the network in questions, including its rhythmic repertoire, stability of generated patterns, etc, and in addition to ability to predict possible transformations before that occur in the system.   
Our greater goal is to gain insight into the fundamental and universal rules governing pattern formation in complex networks of neurons. We believe that one should first investigate the rules underlying the emergence of cooperative rhythms in basic neural motifs, as well as  the role of coupling and in generating a multiplicity of coexisting rhythmic outcomes \cite{krishna2020}.

\section{Acknowledgements}

 We  thank the Brains and Behavior initiative of Georgia State University for the pilot grant support. The authors  thank the past and current lab-mates of the Shilnikov NeurDS lab (Neuro--Dynamical Systems), specifically, K. Wojcik, K. Pusuluri, J. Collens, and J.Schwabedal,  The NeurDS lab is grateful to NVIDIA Corporation for donating the Tesla K40 GPUs that were actively used in this study. This paper was funded in part by the NSF grant IOS-1455527.

\section{Appendix}

The time evolution of the membrane potential, $V$, of each neuron is modeled using the framework of the Hodgkin-Huxley formalism, based on a reduction of a leech heart interneuron model, see \cite{Shilnikov2012} and the references therein:
\begin{equation}
\begin{array}{rcl}
C V^\prime &=& -I_{\mathrm{Na}}-I_{\mathrm{K2}}-I_{\mathrm{L}}-I_{\mathrm{app}}-I_{\rm syn} ,  \\
\tau_{\mathrm{Na}}   h^\prime_{\mathrm{Na}} &=& h^\infty_{\mathrm{Na}}(V)-h, \\
\tau_{\mathrm{K2}}   m^\prime_{\mathrm{K2}} &=& m^\infty_{\mathrm{K2}}(V)-m_{\mathrm{K2}}.
\end{array}\label{eq1}
\end{equation}
The dynamics of the above model involve a fast sodium current, $I_{Na}$  with the activation described by the voltage dependent gating variables, $m_{\mathrm{Na}}$ and $h_{\mathrm{Na}}$, a slow potassium current $I_{K2}$ with the inactivation from
$m_{\mathrm{K2}}$, and an ohmic leak current, $I_\mathrm{leak}$:
\begin{equation}
\begin{array}{rcl}
 I_{\mathrm{Na}}&=&{\bar g}_{\mathrm{Na}}\,m^3_{\mathrm{Na}}\,h_{\mathrm{Na}}\,(V-E_{\mathrm{Na}}),\\
 I_{\mathrm{K2}}&=&{\bar g}_{\mathrm{K2}}\,m_{\mathrm{K2}}^2(V-E_{\mathrm{K}}), \\
 I_{L}&=&\bar g_{\mathrm{L}}\,(V-E_{\mathrm{L}}).
 \end{array}
\end{equation}
$C=0.5\mathrm{nF}$ is the membrane capacitance and $I_{\mathrm{app}}=0.006\mathrm{nA}$ is an applied current.
The values of maximal conductances are ${\mathrm{\bar g}}_{\rm K2}=30\mathrm{nS}$, ${\bar g}_{\rm Na}=160\mathrm{nS}$ and
 $\mathrm{g}_{\rm L}=8\mathrm{nS}$. The reversal potentials are $\mathrm{E}_{\rm Na}=45\mathrm{mV}$, $\mathrm{E}_{\rm K}= -70\mathrm{mV}$ and  $E_{\mathrm{L}}=-46\mathrm{mV}$.  The time constants of gating variables are $\tau_{\rm K2}=0.9\mathrm{s}$ and $\tau_{\rm Na}=0.0405\mathrm{s}$. The steady state values, $h^\infty_{\mathrm{Na}}(V)$, $m^\infty_{\mathrm{Na}}(V)$, $m^\infty_{\mathrm{K2}}(V)$, of the of gating variables are determined by the following Boltzmann equations:
 \begin{equation}
\begin{array}{rcl}
 h^\infty_{\mathrm{Na}}(V)&=&[1+\exp(500(V+0.0325))]^{-1}\\
 m^\infty_{\mathrm{Na}}(V)&=&[1+\exp(-150(V+0.0305))]^{-1}\\
 \quad m^\infty_{\mathrm{K2}}(V)&=&[1+\exp{(-83(V+0.018+\mathrm{V^{shift}_{K2}}))}]^{-1}.
\end{array}
\end{equation}
Fast, non-delayed synaptic currents in this study are modeled using the fast threshold modulation (FTM) paradigm as follows \cite{FTM}:
 \begin{equation}
 \begin{array}{rcl}
I_{\rm syn} &=& g_{\rm syn} (V_{\mathrm{post}}-E_{\rm syn}) \Gamma (V_{\mathrm{pre}}-\Theta_{\rm syn}),\\
\Gamma(V_{\rm pre}-\Theta_{\rm syn}) &=& 1/[1+{\rm exp}\{-1000(V_{\mathrm{pre}}-\Theta_{\rm syn})\}];
\end{array}
\end{equation}
here $V_{\mathrm{post}}$  and $V_{\mathrm{pre}}$ are voltages of the post- and the pre-synaptic cells; the synaptic
threshold $\Theta_{\rm syn}=-0.03\mathrm{V}$ is chosen so that every  spike within a burst in the pre-synaptic cell crosses $\Theta_{\rm syn}$,
see Fig.~1. This implies that the synaptic current, $I_{\rm syn}$, is initiated as soon as $V_{\mathrm{pre}}$ exceeds the
synaptic threshold. The type, inhibitory or excitatory, of the FTM synapse is determined by the level of the reversal potential, $E_{\rm syn}$,
in the post-synaptic cell.
In the inhibitory case, it is set as $E_{\rm syn}=-0.0625\mathrm{V}$ so that $V_{\rm post}(t)> E_{\rm syn}$. In the excitatory case
the level of $E_{\rm syn}$ is raised to zero to guarantee that the average  of $V_{\rm post}(t)$  over the burst period remains below the reversal potential. We point out that alternative synapse models, such as the alpha and other detailed dynamical representation, do not essentially change the dynamical interactions between these cells \cite{pre2012}.

\bibliographystyle{unsrt}

\end{document}